\documentclass[pra,aps,twocolumn,superscriptaddress,showpacs,amsmath,amssymb,floatfix]{revtex4}

\usepackage{graphics}

\begin{document}

\title{Equation of state and effective mass of the unitary Fermi gas 
       \\ in a 1D periodic potential}

\author{Gentaro Watanabe}
\affiliation{CNR INFM-BEC and Department of Physics, University of Trento, 38050 Povo, Italy}
\affiliation{The Institute of Chemical and Physical Research (RIKEN), 2-1 Hirosawa, Wako,
Saitama 351-0198, Japan}
\author{Giuliano Orso}
\affiliation{Laboratoire Physique Th\'eorique et Mod\`eles Statistiques,
Universit\'e Paris Sud, Bat. 100, 91405 Orsay Cedex, France}
\author{Franco Dalfovo}
\affiliation{CNR INFM-BEC and Department of Physics, University of Trento, 38050 Povo, Italy}
\author{Lev P. Pitaevskii}
\affiliation{CNR INFM-BEC and Department of Physics, University of Trento, 38050 Povo, Italy}
\affiliation{Kapitza Institute for Physical Problems, 119334 Moscow, Russia}
\author{Sandro Stringari}
\affiliation{CNR INFM-BEC and Department of Physics, University of Trento, 38050 Povo, Italy}

\date{\today}

\begin{abstract}
By solving the Bogoliubov -- de Gennes equations at zero temperature,
we study the effects of a one-dimensional optical lattice on the behavior 
of a superfluid Fermi gas at unitarity.  We show that, due to the lattice, 
at low densities the gas becomes highly compressible and the effective 
mass is large, with a consequent significant reduction of the sound 
velocity. We discuss the role played by the lattice in the formation of 
molecules and the emergence of two-dimensional effects in the equation of 
state. Predictions for the density profiles and for the frequency of the 
collective oscillations in the presence of harmonic trapping are also given.
\end{abstract}

\maketitle

\section{Introduction}

Ultracold gases in optical lattices provide a new frontier
of research where many remarkable phenomena can be observed and
investigated \cite{BDZ}. By using Feshbach resonances one can tune
the interaction between atoms and investigate the BCS-BEC crossover, 
passing through a resonant regime where the scattering length is very 
large and the system exhibits universal properties (unitary regime) 
\cite{rmp08}. First experiments with Fermi superfluids 
in one-dimensional (1D) optical lattices \cite{Miller} have focused 
on the study of the critical velocity along the crossover and
revealed that superfluidity is particularly robust at unitarity.
The results are in qualitative agreement with the predictions for 
the Landau critical velocity \cite{combescot} and for the behavior 
of the critical current through a single barrier \cite{camerino}.

It is well known that a periodic potential favors the formation of
molecules in a two-component Fermi gas (see \cite{orso} and references
therein), providing an effective shift of the resonance and bound
states in the two-body problem even at unitarity. A major problem is
to understand the consequences of the molecular formation 
in the superfluid phase.  Moreover, for large laser intensities, the 
lattice is expected to give rise to 2D effects.

In order to investigate these properties we use a mean-field theory 
based on the Bogoliubov -- de Gennes (BdG) equations \cite{BdGtheory}.  
Although approximate, this approach captures basic features along 
the whole BCS-BEC crossover \cite{randeria}, including the formation 
of molecules and the most challenging unitary limit where, for uniform 
3D configurations, the predictions are in reasonably good agreement 
with {\it ab initio} Monte Carlo simulations \cite{QMC}. 
The BdG equations apply also to situations where the density varies over 
distances of the order of the healing length. An important example is 
given by configurations with quantized vorticity \cite{sensarma}. 
Furthermore, BdG equations fully account for the modification of 
the scattering properties of fermions induced by the external confinement 
as predicted by Petrov {\it et al.} \cite{petrov} in the limit of a 
deep periodic lattice. As a consequence, for example, a tight 1D lattice 
considerably affects the mean-field superfluid transition temperature, 
where the BCS order parameter vanishes \cite{orso_shlyapnikov}. 
Finally, in the case of a deep lattice the BdG theory is  
expected to approach the 2D mean-field theory of Ref.~\cite{randeria2}.

In this work we study the unitary regime at zero temperature 
focusing on the situation in which the lattice potential is 
relatively weak. In this regime, the tight binding description is not 
adequate and a full numerical approach based on the BdG equations is 
called for. By solving the BdG equations, we first calculate the 
equation of state, the compressibility and the effective mass of the 
unitary Fermi gas in the lattice. The results are then used to obtain 
interesting predictions for observable quantities such as the sound 
velocity, the frequency of collective modes, and the density profile
of trapped gases in typical experimental configurations.

\section{Formalism}

At zero temperature, the chemical potential $\mu$ of a superfluid Fermi gas 
in a lattice is given by the derivative of the energy density $e=E/V$ 
with respect to the average (coarse-grained) density $n$:
\begin{equation}
  \mu = \frac{ \partial e(n,P)}{ \partial n}\, ,
\label{mu}
\end{equation}
where $P$ is the quasi-momentum of the superfluid along the lattice 
\cite{current}.  The compressibility $\kappa$ and the effective mass $m^*$,
are given by the second derivatives of $e$ with respect to $n$ and $P$:
\begin{equation}
  \kappa^{-1} = n\frac{\partial^2 e(n,P)}{\partial n^2} = 
n \frac{\partial \mu(n,P) }{ \partial n} \ ; \quad  
\frac{1}{m^*} = \frac{1}{n} \frac{\partial^2 e(n,P)}{\partial P^2}\, . 
\label{kappa_m*}
\end{equation}
We calculate these quantities at unitarity for $P=0$, i.e. for a gas at 
rest, in the periodic potential 
\begin{equation}
  V_{\rm ext}(z)=sE_{\rm R} \sin^2q_{\rm B}z .
\label{lattice}
\end{equation}
Here $s$ is the laser intensity, $E_{\rm
R}=\hbar^2q_{\rm B}^2/2m$ is the recoil energy, $q_{\rm B}=\pi/d$ is
the Bragg wave vector, $d$ is the lattice constant and $m$ is the atom 
mass. Then the BdG equations are given by
\begin{equation}
\left( \begin{array}{cc}
H'(\mathbf r) & \Delta (\mathbf r) \\
\Delta^\ast(\mathbf r) & -H'(\mathbf r) \end{array} \right)
\left( \begin{array}{c} u_i( \mathbf r) \\ v_i(\mathbf r)
\end{array} \right)
=\epsilon_i\left( \begin{array}{c} u_i(\mathbf r) \\
v_i(\mathbf r) \end{array} \right) \;,
\label{BdGnonuniform}
\end{equation}
where $H'(\mathbf r) =-\hbar^2 \nabla^2/2m +V_{\rm ext}-\mu$.
The order parameter $\Delta (\mathbf r) $ and the chemical potential 
$\mu$, appearing in Eq.~(\ref{BdGnonuniform}), are variational 
parameters determined from the self-consistency relation
\begin{equation}\label{gap}
\Delta(\mathbf r) =-g \sum_i u_i(\mathbf r) v_i^*(\mathbf r) ,
\end{equation}
together with the constraint 
$n=(2/V)\sum_i \int \left| v_i(\mathbf r) \right|^2d{\bf r}$, enforcing 
conservation of the average density $n$. In Eq.~({\ref{gap}), $g$ is the 
coupling constant for the contact interaction and the BdG eigenfunctions 
obey the normalization condition $\int d{\bf r} \left[u_i^*({\bf r})u_j({\bf
r}) +v_i^*({\bf r})v_j({\bf r})\right]=\delta_{i,j}$. 

For contact interactions, the right hand side of Eq.~(\ref{gap}) is
ultraviolet divergent and must be cured by the pseudo-potential method. 
A standard procedure consists of introducing a cut-off energy 
$E_{\rm C}$ in the sums over the BdG eigenstates and of replacing 
the bare coupling constant by the $s$-wave scattering length $a_{\rm s}$ 
through the relation $(k_{\rm F}a_{\rm s})^{-1} =8\pi E_{\rm F}/(gk_{\rm
F}^3)+(2/\pi)\sqrt{E_{\rm C}/E_{\rm F}}$, where $k_F= (3\pi^2
n)^{1/3}$ and $E_{\rm F}= \hbar^2 k_F^2/(2m)$ are the Fermi wave vector
and energy, respectively, of a uniform noninteracting Fermi gas with
density $n$. An alternative regularization scheme is the one proposed in 
Refs.~\cite{bruun,bulgac}. This method exploits directly the short 
distance behavior of the single-particle Green's function and is efficient 
even in the presence of a tight confinement. Both procedures give identical
results provided $E_{\rm C}$ is large enough.

In the presence of a supercurrent with wave vector $Q=P/\hbar$
moving along the lattice, one can write the order parameter in the form
$\Delta(\mathbf r)=e^{i 2Q z}\tilde{\Delta}(z)$, where
$\tilde{\Delta}(z)$ is a complex function with period $d$. 
Therefore, from Eq.~(\ref{gap}), we see that the eigenfunctions of 
Eq.~(\ref{BdGnonuniform}) must have the Bloch form $u_i(\mathbf r) =
\tilde{u}_i(z) e^{i Q z}e^{i\mathbf k \cdot \mathbf r }$ and
$v_i(\mathbf r) = \tilde{v}_i(z) e^{-i Q z}e^{i\mathbf k \cdot
\mathbf r }$, where $k_z$ lies in the first Brillouin zone and
$\tilde{u}_i$ and $\tilde{v}_i$ are periodic in $z$ with period $d$. 
This transformation reduces Eq.~(\ref{BdGnonuniform}) to 
the BdG equations for $\tilde{u}_i$ and $\tilde{v}_i$ as
\begin{equation}
\left( \begin{array}{cc}
\tilde{H}'_{Q}(z) & \tilde{\Delta}(z) \\
\tilde{\Delta}^\ast(z) & -\tilde{H}'_{-Q}(z) \end{array} \right)
\left( \begin{array}{c} \tilde{u}_i(z) \\ \tilde{v}_i(z)
\end{array} \right)
=\epsilon_i\left( \begin{array}{c} \tilde{u}_i(z) \\
\tilde{v}_i(z) \end{array} \right) \;,
\label{BdGnonuniform2}
\end{equation}
where
\begin{equation}
  \tilde{H}'_{Q}(z)\equiv \frac{\hbar^2}{2m} \left[k^2_x+k^2_y
+\left(-i\partial_z+Q+k_z\right)^2\right]+V_{\rm ext}(z) -\mu\, .
\label{hq}
\end{equation}
From now on, 
the label $i$ represents the wave vector $\mathbf k$ as well as the 
band index.

 From the solution of the BdG equations, we can directly calculate
$\mu$ and $\kappa$ [see Eqs.~(\ref{mu}) and (\ref{kappa_m*})], 
while $m^*$ is obtained from the energy density $e(n,P)$ \cite{BdGtheory}
\begin{equation}
e = \int \! d{\bf r}\left[\sum_i 2(\mu-\epsilon_i )|\tilde{v}_i(z)|^2
+\sum_i \tilde\Delta^*(z) \tilde{u}_i(z)\tilde{v}_i^*(z) \right] .
\label{E}
\end{equation}
Note that in the rhs of Eq.~(\ref{E}) both contributions are
separately divergent but the sum is finite, as one can easily check
for the uniform case $(s=0)$.

\section{Results}

A first important remark concerns the low density limit of $\mu$ and 
$m^*$ whose values are found in perfect agreement with the 
results of the exact solution of the two-body problem \cite{orso}.  
This proves that the BdG theory correctly accounts for the deep 
modifications of the atomic scattering properties induced by the 
external confinement which gives rise to bound molecules even at 
unitarity. 

Our results for the density dependence of $\kappa^{-1}$ and $m^*$ 
at unitarity are shown in Figs.\ \ref{fig_kinv_meff_uni}(a) and 
\ref{fig_kinv_meff_uni}(b). They are plotted for $s=1$, 2.5, and 5, 
as functions of $E_{\rm F}/E_{\rm R}= (k_{\rm F}d/\pi)^2$.
The solid, dashed and dashed-dotted lines correspond to the values 
obtained with the regularization scheme of Ref.~\cite{bulgac} and 
choosing the cut-off energy large enough to ensure
convergence  within $1$-$2$\% \cite{cutoff}. In order to appreciate 
the role of the cut-off energy, we also plot the results obtained using
the standard regularization procedure with $E_{\rm C}=100 E_{\rm F}$
(points). The results agree within $10$-$15$\%, the accuracy being
worse in the small density regime $E_{\rm F} \ll E_{\rm R}$.

In the absence of the lattice ($s=0$) the results take a universal
behavior at unitarity: the only relevant length being the
interparticle distance fixed by $k_{\rm F}$. Due to translational
invariance, one can write $e(n,P) = e(n,0) + n P^2/2m$ so that
$m^*=m$. Furthermore, the energy density at $P=0$ can be written as
$e(n,0)=(1+\beta)e^0(n,0)$, where $e^0(n,0)\equiv (3/5)n E_{\rm F}$ is
the ideal Fermi gas value and $\beta$ is the dimensionless universal
parameter accounting for the interactions in uniform gas. The BdG
equations predict $\beta\simeq -0.41$ to be compared with the Monte 
Carlo result $\beta\simeq -0.58$ \cite{QMC}.

\begin{figure}[htbp]
\begin{center}\vspace{0.0cm}
\rotatebox{0}{
\resizebox{8.1cm}{!}{\includegraphics{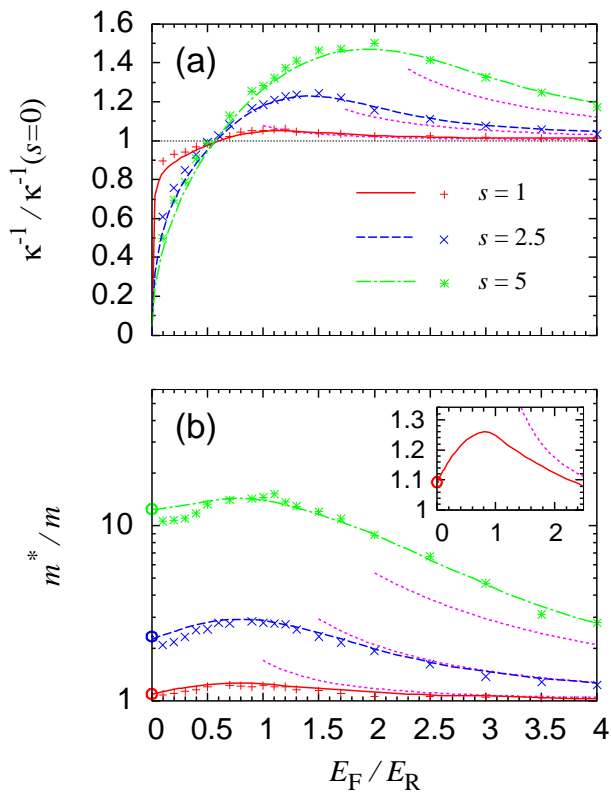}}}
\caption{\label{fig_kinv_meff_uni}(Color online)\quad 
Inverse compressibility $\kappa^{-1}$, and effective mass $m^*$ of 
the unitary Fermi gas for $s=1$ (red), $2.5$ (blue), and $5$ (green). 
Lines are convergent results obtained by the regularization
scheme of Ref.~\cite{bulgac}. Points are obtained by means of
the standard regularization scheme with $E_{\rm C}=100 E_{\rm F}$.
Asymptotic expressions (\ref{expansionk-1}) and (\ref{expansionm}) 
are shown by the dotted lines. 
Open circles in panel (b) show $m^*$ obtained in Ref.~\cite{orso}.
The $s=1$ results for $m^*$ are also shown in the inset in the 
linear scale. 
}
\end{center}
\end{figure}

\subsection{Unitary Fermi gas in a lattice}

\subsubsection{Equation of State}

New features appear when $s$ increases. Let us first
discuss the behavior of the equation of state. At small densities 
($E_{\rm F}/E_{\rm R} \ll 1$) we find that the lattice causes 
a linear density dependence of the chemical potential [see inset 
of Fig.\ \ref{fig_densprof}] and a strong suppression of 
$\kappa^{-1}$ with respect to the uniform value [see Fig.\
\ref{fig_kinv_meff_uni}(a)]. These results are consistent with  
the tendency of the lattice to favor the formation of molecules. 
The size of these molecules is fixed by the values of $s$ and $d$ so 
that, in the limit of a dilute gas ($k_{\rm F}d \ll 1$), the interparticle 
distance can be larger than the molecular size. In this limit, the BdG 
equation describes the formation of a molecular gas which gives rise to  
Bose-Einstein condensation; the equation of state is given by the 
Bogoliubov theory and $\kappa^{-1}$ is expected to be linear in the 
density, yielding $\kappa^{-1}/\kappa^{-1}(s=0) \propto n^{1/3}
\rightarrow 0$. For $s \gg 1$ the chemical potential remains 
almost linear in density even at relatively large densities, due 
to 2D effects caused by the bandgap in the longitudinal motion (see 
discussion below). At even higher densities, one eventually recovers 
the behavior of a uniform gas since the lattice only provides a small
perturbation. By using an hydrodynamic approach and expanding in the
small parameter $sE_{\rm R}/E_{\rm F}$, we find \cite{note_hydro}
\begin{eqnarray}
\kappa^{-1} \simeq \frac{2}{3}(1+\beta)
E_{\rm F} \left[ 1 +  \frac{1}{32} (1+\beta)^{-2} 
\left(\frac{sE_{\rm R}}{E_{\rm F}}\right)^2 \right] \nonumber \\
 + O \left[ \left( sE_{\rm R}/E_{\rm F}\right)^4 \right]
\label{expansionk-1}
\end{eqnarray}
This is shown by dotted lines in Fig.~\ref{fig_kinv_meff_uni}(a). 

\begin{figure}[htbp]
\begin{center}\vspace{0.0cm}
\rotatebox{0}{
\resizebox{8cm}{!}
{\includegraphics{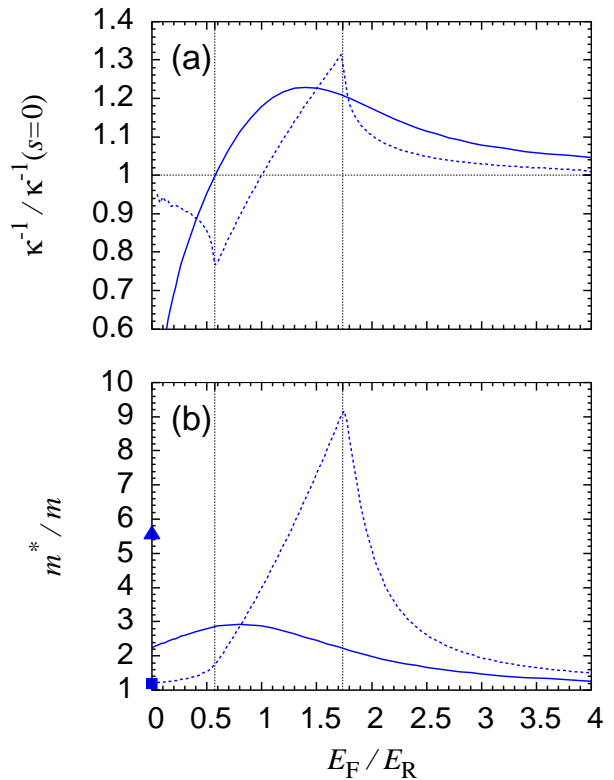}}}
\caption{\label{fig_comp}(Color online)\quad Comparison between the
unitary Fermi gas (solid lines) and the ideal Fermi gas (dashed lines)
in the same optical lattice with $s=2.5$. The region between the two
vertical lines represents the range where the chemical potential of
the ideal gas lies in the bandgap above the lowest Bloch band. In
panel (b), the square at $m^*/m=1.20$ shows the value of the effective
mass of a single atom with mass $m$ in the same lattice, while the
triangle at $m^*/m=5.53$ shows that of a point-like molecule with mass
$2m$.}
\end{center}
\end{figure}

\subsubsection{Effective Mass}

The formation of molecules has important consequences also for $m^*$. 
Due to this effect, at low densities, the enhancement of $m^*$ 
caused by lattice is much more dramatic in the unitary Fermi gas 
compared to the ideal Fermi gas (or, equivalently to $m^*$ in the 
BCS limit) and to the gas of bosons with the same mass $m$ 
(see Fig.\ 4 in Ref.~\cite{boson}).
We also note that, for a given $s$, the value 
of $m^*$ at $E_{\rm F}\rightarrow 0$ lies between the value of $m^*$ 
for a single atom of mass $m$ in the same lattice [square in the vertical 
axis of Figs.\ \ref{fig_comp}(b)] and the value of $m^*$ calculated for 
a point-like molecule of mass $2m$ (triangle).
As $E_{\rm F}/E_{\rm R}$ increases, the effective mass exhibits a
maximum at $E_{\rm F}/E_{\rm R}\sim 1$, 
then it decreases towards the bare mass, $m^*=m$. 
The same hydrodynamic argument used for
$\kappa^{-1}$ also explains the 
behavior of the effective mass for small $sE_{\rm R}/E_{\rm F}$:
\begin{equation}
\frac{m^*}{m} \simeq 1+\frac{9}{32} (1+\beta)^{-2} 
\left(\frac{s E_{\rm R}}{E_{\rm F}}\right)^2 
+ O\left[\left(sE_{\rm R}/E_{\rm F}\right)^4\right]\ ,
\label{expansionm}
\end{equation}
showing that the effect of the lattice is stronger for $m^*$ than for
$\kappa^{-1}$.  It is worth comparing the results with the case
of bosonic atoms, where $m^*$ decreases monotonically with increasing
density since the interaction broadens the condensate wave function
and favors the tunneling \cite{boson}.

\subsubsection{Comparison with the Ideal Fermi Gas}

The occurrence of a maximum in the curves for both $\kappa^{-1}$ and
$m^*$ can be interpreted as an effect of the energy gap in the
longitudinal motion, which opens at $q=q_{\rm B}$.  An instructive
comparison can be made with an ideal Fermi gas in the same lattice, 
where the effects of the bandgap are more evident due to the sharper 
Fermi surface.
In Fig.\ \ref{fig_comp} one
sees that the ideal gas curves have two cusps. They occur precisely when
$\mu$ coincides with the top of the lowest band and the bottom of the
first excited band at $q=q_{\rm B}$, respectively. In between, as
$E_{\rm F}/E_{\rm R}$ increases, only the transverse modes are
available and the system behaves effectively like a 2D system,
the longitudinal degree of freedom simply giving a
constant contribution to $\kappa^{-1}$ and $m^*$. Consequently,
$\kappa^{-1}$ and $m^*$ are proportional to the average density in
this region (see Eqs.~(10) and (11) in Ref.~\cite{pso}) and one
finds the power laws $\kappa^{-1}/\kappa^{-1}(s=0) \propto E_{\rm
F}^{1/2}$ and $m^* \propto E_{\rm F}^{3/2}$. The interval
where this 2D behavior takes place becomes wider as $s$ increases,
since the bandgap increases with $s$.
By comparing the ideal
and unitary Fermi gases, we see that the interaction significantly
smears the effect of the gap as a result of a much broader Fermi
surface. Especially the maximum values of $m^*$ are drastically
reduced. Note also that the molecular-like pair correlations, which
are responsible for the low density behavior of $\kappa^{-1}$ and
$m^*$, are absent in the ideal Fermi gas.

\begin{figure}[htbp]
\begin{center}\vspace{0.0cm}
\rotatebox{0}{
\resizebox{8cm}{!}
{\includegraphics{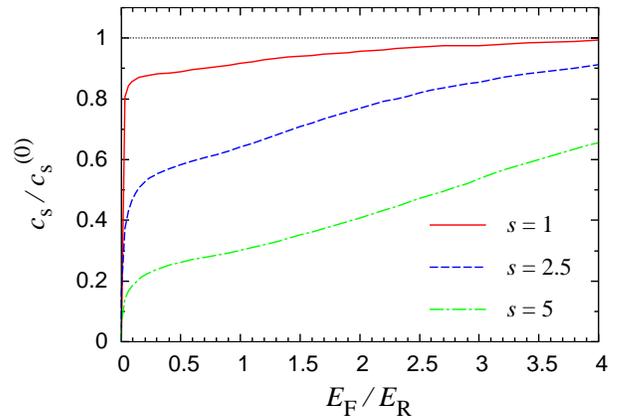}}}
\caption{\label{fig_cs}(Color online)\quad
Sound velocity $c_{\rm s}$ in units of the sound velocity
$c_{\rm s}^{(0)} =[(2/3)(1+\beta)E_{\rm F}/m]^{1/2}$
for the uniform system. As in Fig.\ \ref{fig_kinv_meff_uni},
red, blue and green lines correspond to $s=1, 2.5$, and $5$,
respectively.
}
\end{center}
\end{figure}

\subsubsection{Sound Velocity}

Our results for $\kappa^{-1}$ and $m^*$ can be used to calculate the
sound velocity $c_{\rm s}=\sqrt{\kappa^{-1}/m^*}$, which is given in 
Fig.~\ref{fig_cs}. One can see a significant
reduction of $c_{\rm s}$ compared to the uniform system.  Except for
very low densities, the sound velocity varies rather smoothly with 
the density.  This makes it possible to provide an estimate the 
change of $c_{\rm s}$ induced by the lattice even for harmonically 
trapped gases, where the coarse-grained density is not uniform. 
Notice that the propagation of sound is a direct consequence
of superfluidity and, in the presence of tight lattices, can be
regarded as a Josephson effect, where the gas tunnels in a coherent
way through the barriers produced by the lattice.

\subsection{Unitary Fermi gas in a lattice + trap}

\subsubsection{Density Profile}

We can also provide useful predictions for the
density profile and for the collective motion of a trapped gas, when a
harmonic confinement is added to the periodic potential. The coarse-grained 
density profile, $n(z)$, is easily calculated using the local
density approximation (LDA) for $\mu$. Figure \ref{fig_densprof} clearly
shows that, for $s=5$, the profile takes the form of an inverted
parabola, reflecting the linear density dependence of the chemical
potential (see inset).  In this calculation, we set
$\omega_{\perp}=\omega_z$, where $\omega_\perp$ and $\omega_z$ are the
transverse and longitudinal trapping frequencies,
$\hbar\omega_z/E_{\rm R}=0.01$, and the number of particles $N=10^6$;
these parameters are close to the experimental ones in Ref.~\cite{Miller}.

The accurate measurement of the density profile in the presence of
harmonic trapping can actually give direct information on the
compressibility of the gas. In fact, within the range of validity 
of LDA, the two quantities obey a simple relationship which, for
isotropic traps, is 
\begin{equation}
\frac{\partial n}{\partial r} = -m \omega^2r 
\left(\frac{\partial \mu}{\partial n}\right)^{-1} .
\end{equation}

\begin{figure}[htbp]
\begin{center}\vspace{0.0cm}
\rotatebox{0}{
\resizebox{8cm}{!}
{\includegraphics{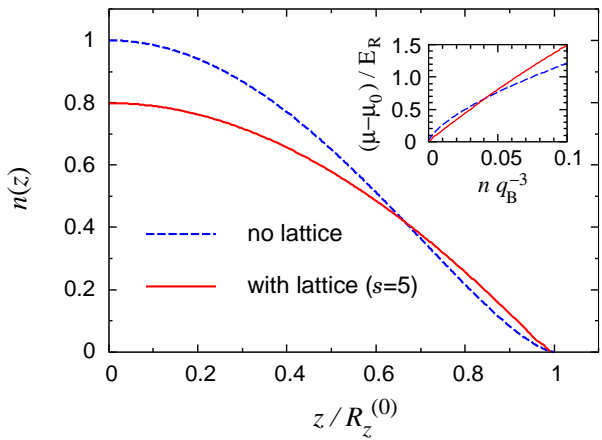}}}
\caption{\label{fig_densprof}(Color online)\quad Coarse-grained
density profiles of a trapped gas, $n(r_{\perp}=0,z)$ for $s=0$
and $5$ in units of the central density $n(0)=0.0869 q_{\rm B}^3$
calculated for $s=0$ (this local density corresponds to
$E_{\rm F}/E_{\rm R}=1.88$). The quantity $R_z^{(0)}$ is the axial
Thomas-Fermi radius for $s=0$. The inset shows the density
dependence of the chemical potential. The parameters of the
trap are given in the text.
}
\end{center}
\end{figure}

\subsubsection{Collective Modes}

The collective modes can be studied by solving the hydrodynamic
equations, where the periodic potential is included through its 
effects on the equation of state $\mu(n)$ and on the effective
mass which determines the current in the longitudinal direction
\cite{book}. While the frequency of the dipole (center-of-mass) 
oscillation along the transverse direction is not affected by the 
lattice, a useful estimate for the dipole frequency in the 
$z$-direction can be obtained using a sum rule approach 
based on the calculation of the energy weighted and inverse energy 
weighted moments of the dipole strength within the hydrodynamic 
theory \cite{pso}. This approach yields the
result $\omega_D=\omega_z (m /\, \overline{m^*})^{1/2}$ where
\begin{equation}
\frac{1}{\ \overline{m^*}\ }
= \frac{1}{N} \int d{\bf r}\ \frac{n({\bf r})}{m^*[n({\bf r})]}\ .
\label{sumrule}
\end{equation}
Even without calculating $n({\bf r})$, 
we can easily estimate lower and upper bounds for the frequency by
replacing $\overline{m^*}$ with the maximum and minimum values of
$m^*$ as a function of $n$ as one moves from the center to the border
of the atomic cloud. For clouds whose maximum density is such that
$E_{\rm F}/E_{\rm R}\alt 1.5$ the minimum value corresponds to the low
density $E_{\rm F}/E_{\rm R}\to 0$ limit, while the maximum corresponds
to the absolute maximum of the curves in Fig.\
\ref{fig_kinv_meff_uni}(b). In this way we obtain the estimate $0.89
\le \omega_D/\omega_z \le 0.96$ for $s=1$, $0.59\le \omega_D/\omega_z
\le 0.66$ for $s=2.5$, and $0.26\le \omega_D/\omega_z \le 0.28$ for
$s=5$.

Finally, the effects of the lattice on the equation
of states can be observed also by studying the compression modes.
For example, taking the cigar shape geometry $\omega_z\ll
\omega_{\perp}$, the frequency of the radial breathing mode  is 
unaffected by $m^*$. In the regime where the chemical potential is linear 
in the density (see inset of Fig.\ \ref{fig_densprof}) the collective
frequency approaches the value $\omega=2 \omega_\perp$, while its
value is $\sqrt{10/3}\, \omega_\perp$ in the absence of the
lattice. The transition between the two different regimes is similar
to the one recently investigated by varying the
scattering length on the BEC side of the resonance in the absence of
periodic potentials (see, e.g., \cite{rmp08}).

\section{Conclusion}

In this work we have shown that the inclusion of a 1D optical
lattice, by favoring the formation of molecular configurations and
by inducing a band structure in the quasiparticle spectrum,
has profound consequences on the thermodynamic quantities, 
the density profile, and the collective oscillations 
of the unitary Fermi gas.
Our calculations can be naturally 
extended to the case of finite quasi-momenta, where energetic and 
dynamic instabilities are expected to impose some limits to the 
superfluid motion. Further investigations of the 2D nature
of the many-body system achievable with large laser intensities are
also in progress. 

\acknowledgments

We thank M. Antezza, G. Bruun,
E. Furlan, S. Giorgini, Y. Ohashi, and M. Urban
for fruitful discussions.
G.\,O. is supported by the Marie Curie Fellowship under contract 
n. EDUG-038970.
This work has been supported by MIUR and by Fermix-Euroquam.


\begin{thebibliography}{99}
\bibitem{cutoff} In practice, we include more than $20$ single-particle 
Bloch bands in these calculations.
\bibitem{note_hydro} In the limit of $E_{\rm F}\gg E_{\rm R}$, the 
length scale of the density variation $\sim d$ is much larger than 
the coherence length $\xi \sim k_{\rm F}^{-1}$ (at unitarity) since 
$(d/\xi)^2\sim (k_{\rm F}/q_{\rm B})^2 \gg 1$, 
which validates the hydrodynamic theory.
\end{thebibliography}
\end{document}